\documentclass[aps,prc,preprint,showpacs,showkeys]{revtex4}
\usepackage{amssymb}
\usepackage{amsmath}
\usepackage{graphicx}
\usepackage{bm}
\usepackage{epsf}

\begin{document}

\title{Symmetry energy of hot nuclei in the relativistic Thomas-Fermi approximation}

\author{Z. W. Zhang}
\affiliation{School of Physics, Nankai University, Tianjin 300071, China}
\author{S. S. Bao}
\affiliation{School of Physics, Nankai University, Tianjin 300071, China}
\author{J. N. Hu}
\affiliation{School of Physics, Nankai University, Tianjin 300071, China}
\author{H. Shen}~\email{shennankai@gmail.com}
\affiliation{School of Physics, Nankai University, Tianjin 300071, China}

\begin{abstract}
We develop a self-consistent description of hot nuclei within
the relativistic Thomas--Fermi approximation using the
relativistic mean-field model for nuclear interactions.
The temperature dependence of the symmetry energy and other
physical quantities of a nucleus are calculated by employing
the subtraction procedure in order to isolate the nucleus
from the surrounding nucleon gas. It is found that the
symmetry energy coefficient of finite nuclei is
significantly affected by the Coulomb polarization effect.
We also examine the dependence of the results on nuclear
interactions and make a comparison between the results
obtained from relativistic and nonrelativistic
Thomas-Fermi calculations.
\end{abstract}

\pacs{21.10.Dr, 21.30.Fe, 21.65.Ef}
\keywords{Symmetry energy, Thomas-Fermi approximation, Relativistic mean-field model}
\maketitle

% 21.65.Ef Symmetry energy
% 21.10.Dr Binding energies and masses
% 21.30.Fe Forces in hadronic systems and effective interactions

%\newpage
%%%%%%%%%%%%%%%%%%%%%%%%%%%%%%%%%%%%%%%%%%%%%%%%%%%%%%%%%%%%%%%%%%%%%
\section{Introduction}
\label{sec:1}

The properties of hot nuclei, such as excitation energies, entropies,
symmetry energies, and density distributions, are of great interest
in both nuclear physics and astrophysics~\cite{RPP05}.
Especially, the symmetry energy and its dependence on density
and temperature play a crucial role in understanding various
phenomena in heavy-ion collisions, supernova explosions,
and neutron-star properties~\cite{PR05,PR07,LiBA08}.
Hot nuclei formed in nucleus-nucleus collisions are
thermodynamically unstable against the emission of nucleons.
Theoretically, an external pressure is imposed on the nucleus
to compensate for the tendency of nucleon emission.
This pressure is assumed to be exerted by a surrounding gas
representing evaporated nucleons, which is in equilibrium
with the hot nucleus.
In order to isolate the nucleus from the surrounding gas,
a subtraction procedure was first proposed
in the Hartree-Fock framework~\cite{BLV85}
and then used in the Thomas-Fermi approach~\cite{TF87}.
The subtraction procedure is based on the existence of
two solutions to the equations of motion, one corresponding to
the nuclear liquid phase in equilibrium with the surrounding gas ($NG$)
and the other corresponding to the nucleon gas alone ($G$).
The density profile of the nucleus ($N$) is then obtained by
subtracting the gas density from that of the liquid-plus-gas
phase. As a result, physical quantities of the isolated nucleus
obtained using the subtraction procedure could be independent of
the size of the box in which the calculation is performed.
This subtraction procedure has been widely used
in the nonrelativistic Thomas-Fermi approximation with Skyrme
effective interactions~\cite{Sub01,Sub02,Sub07,Sub12,PLB12,Sub14,Sub14b}.
It is interesting and important to develop a relativistic
Thomas-Fermi model for the description of hot nuclei by 
employing the subtraction procedure and to investigate
the temperature dependence of the symmetry energy
of finite nuclei.

The relativistic Thomas-Fermi approximation has been developed
and applied to study the droplet formation~\cite{RTF99a,RTF99b}
and nuclear pasta phases~\cite{Mene08,Mene10,Gril12}
in asymmetric nuclear matter at subnuclear densities.
This method is considered to be self-consistent in
the treatment of surface effects and nucleon distributions.
The relativistic Thomas-Fermi approximation has been
used to describe finite nuclei~\cite{RTF02,RTF01}
and nonuniform nuclear matter for supernova simulations~\cite{Zhang14}.
In Refs.~\cite{RTF02,RTF01}, the caloric curve for finite nuclei
was studied within the relativistic Thomas-Fermi approximation,
and the results obtained were found to depend on the input
freeze-out volume, which is actually the box size for
performing the calculation.
In the present work, we aim to study the properties of hot nuclei
within the relativistic Thomas--Fermi approximation by employing the
subtraction procedure, so that the results obtained can be
independent of the box size.
For the nuclear interaction, we adopt the relativistic mean-field (RMF)
model, which has been successfully used to study various
phenomena in nuclear physics~\cite{Sero86,Ring90,Meng06}.
In the RMF approach, nucleons interact via the exchange of
scalar and vector mesons, and the parameters
are generally fitted to nuclear matter saturation properties
or ground-state properties of finite nuclei.
In the present calculations, we consider four different
RMF parametrizations, NL3~\cite{NL3}, TM1~\cite{TM1},
FSU~\cite{FSU}, and IUFSU~\cite{IUFSU}, so that we can examine
the dependence of results on the RMF parametrization.
These RMF parametrizations are known to be successful in describing
the ground state properties of finite nuclei including unstable ones.
In this work, we intend to adopt these different RMF parametrizations
to study the properties of hot nuclei and investigate the temperature
dependence of the symmetry energy of finite nuclei
within the relativistic Thomas-Fermi approximation.

Because of the increasing importance of the symmetry energy
in nuclear physics and astrophysics, there have been numerous
studies on the symmetry energy of nuclear matter
based on various many-body methods~\cite{PR05,PR07,LiBA08}.
However, there are fewer calculations for the symmetry energy
of finite nuclei~\cite{SEA03,SEA09,SEA02,SEA10,Oyam10,Yao12,Dong13}.
In Ref.~\cite{SEA10}, the symmetry energy of finite
nuclei was calculated based on a density-functional theory
at both zero and finite temperatures.
A parametrized Thomas-Fermi approach was used in Ref.~\cite{Oyam10}
to estimate the density dependence of the symmetry energy
from nuclear masses. In Ref.~\cite{Dong13}, the symmetry energy
coefficients of finite nuclei were extracted in the framework
of the Skyrme-Hartree-Fock approach.
The nonrelativistic Thomas-Fermi approximation with Skyrme
effective interactions was employed for evaluating the
symmetry energy of finite nuclei and its dependence
on temperature~\cite{Sub12,PLB12}.
In a recent work by Agrawal {\it et al.} \cite{Sub14b},
various definitions of the symmetry energy coefficients
of infinite nuclear matter and finite nuclei,
as well as their temperature dependencies, have been
extensively investigated by using Skyrme interactions,
and it was found that the symmetry energy of nuclear
matter has a weak dependence on temperature, while that of
finite nuclei shows a rapid decrease with increasing temperature.
In the present work, we will use the relativistic Thomas-Fermi
approximation with the RMF model for nuclear interactions
to study the temperature dependence of the symmetry energy
of finite nuclei. We will also make a comparison
between the relativistic and nonrelativistic results.

This article is organized as follows. In Sec.~\ref{sec:2},
we briefly describe the RMF model and the relativistic
Thomas-Fermi approximation by employing the subtraction
procedure for the description of hot nuclei.
In Sec.~\ref{sec:3}, we present the numerical results of
hot nuclei properties and the temperature dependence of
the symmetry energy of finite nuclei.
Section~\ref{sec:4} is devoted to the conclusions.

%%%%%%%%%%%%%%%%%%%%%%%%%%%%%%%%%%%%%%%%%%%%%%%%%%%%%%%%%%%%%%%%%%%%%
\section{Formalism}
\label{sec:2}

In this section, we first give a brief description of the RMF model
used for nuclear interactions.
Then we develop a relativistic Thomas-Fermi model by employing the
subtraction procedure for the description of hot nuclei.
The symmetry energy of finite nuclei can be calculated
by a pair of nuclei that have the same mass number ($A$) but
different numbers of protons ($Z$) and neutrons ($N$).

\subsection{Relativistic mean-field model}
\label{sec:2.1}

In the RMF approach, nucleons interact through the exchange of various mesons.
The mesons considered are the isoscalar scalar and vector mesons
($\sigma$ and $\omega$) and isovector vector meson ($\rho$).
The nucleonic Lagrangian density reads
\begin{eqnarray}
\label{eq:LRMF}
\mathcal{L}_{\rm{RMF}} & = & \sum_{i=p,n}\bar{\psi}_i
\left[i\gamma_{\mu}\partial^{\mu}-M-g_{\sigma}\sigma
      -g_{\omega}\gamma_{\mu}\omega^{\mu}-\frac{g_{\rho}}{2}\gamma_{\mu}\tau_a\rho^{a\mu}
      \right]\psi_i  \notag \\
&& +\frac{1}{2}\partial_{\mu}\sigma\partial^{\mu}\sigma
   -\frac{1}{2}m^2_{\sigma}\sigma^2-\frac{1}{3}g_{2}\sigma^{3}
   -\frac{1}{4}g_{3}\sigma^{4} \notag \\
&& -\frac{1}{4}W_{\mu\nu}W^{\mu\nu} +\frac{1}{2}m^2_{\omega}\omega_{\mu}\omega^{\mu}
   +\frac{1}{4}c_{3}\left(\omega_{\mu}\omega^{\mu}\right)^2  \notag \\
&& -\frac{1}{4}R^a_{\mu\nu}R^{a\mu\nu} +\frac{1}{2}m^2_{\rho}\rho^a_{\mu}\rho^{a\mu}
   +\Lambda_{\rm{v}} \left(g_{\omega}^2\omega_{\mu}\omega^{\mu}\right)
    \left(g_{\rho}^2\rho^a_{\mu}\rho^{a\mu}\right),
\end{eqnarray}
where $W^{\mu\nu}$ and $R^{a\mu\nu}$ are the antisymmetric
field tensors for $\omega^{\mu}$ and $\rho^{a\mu}$, respectively.
In the RMF approach, meson fields are treated as
classical fields and the field operators are replaced by their expectation
values. For a static system, the nonvanishing expectation values are
$\sigma =\left\langle \sigma \right\rangle$, $\omega =\left\langle
\omega^{0}\right\rangle$, and $\rho =\left\langle \rho^{30} \right\rangle$.

For nonuniform nuclear matter at finite temperature, the local energy
density derived from the Lagrangian density~(\ref{eq:LRMF}) is given by
\begin{eqnarray}
\label{eq:ermf}
{\varepsilon}(r) &=&\displaystyle{\sum_{i=p,n}
\frac{1}{\pi^{2}}\int_{0}^{\infty}dk\,k^{2}\,\sqrt{k^{2}+{M^{\ast }}^{2}}
\left( f_{i+}^{k}+f_{i-}^{k}\right) }  \nonumber \\
&&+\frac{1}{2}(\nabla \sigma )^{2}+\frac{1}{2}m_{\sigma }^{2}\sigma ^{2}+%
\frac{1}{3}g_{2}\sigma ^{3}+\frac{1}{4}g_{3}\sigma ^{4}  \nonumber \\
&&-\frac{1}{2}(\nabla \omega )^{2}-\frac{1}{2}m_{\omega }^{2}\omega ^{2}-%
\frac{1}{4}c_{3}\omega ^{4}+g_{\omega }\omega \left( n_{p}+n_{n}\right)
\nonumber \\
&&-\frac{1}{2}(\nabla \rho )^{2}-\frac{1}{2}m_{\rho }^{2}\rho ^{2}+\frac{%
g_{\rho }}{2}\rho \left( n_{p}-n_{n}\right)  \nonumber \\
&&-{\Lambda }_{\text{v}}\left( g_{\omega }^{2}{\omega }^{2}\right) \left(
g_{\rho }^{2}{\rho }^{2}\right) ,
\end{eqnarray}
where $M^{\ast}=M+g_{\sigma}\sigma$ is the effective nucleon mass
and $n_{i}$ is the number density of species $i$ ($i=p$ or $n$).
The entropy density is given by
\begin{eqnarray}
\label{eq:srmf}
s(r)=\displaystyle{\sum_{i=p,n}\frac{1}{\pi^{2}} \int_{0}^{\infty}dk\,k^{2}}
& \left[ -f_{i+}^{k}\ln f_{i+}^{k}-\left( 1-f_{i+}^{k}\right) \ln \left(
1-f_{i+}^{k}\right) \right.  \nonumber \\
& \left. -f_{i-}^{k}\ln f_{i-}^{k}
  -\left( 1-f_{i-}^{k}\right) \ln \left( 1-f_{i-}^{k}\right) \right],
\end{eqnarray}
where $f_{i+}^{k}$ and $f_{i-}^{k}$ ($i=p$ or $n$) are the occupation probabilities
of the particle and antiparticle at momentum $k$, respectively.
The number density of protons ($i=p$) or neutrons ($i=n$) at position $r$ is given by
\begin{equation}
\label{eq:nirmf}
 n_{i}(r)=\frac{1}{\pi^2}
       \int_0^{\infty} dk\,k^2\,\left(f_{i+}^{k}-f_{i-}^{k}\right).
\end{equation}

In the RMF model, the parameters are generally fitted to nuclear matter
saturation properties or ground-state properties of finite nuclei.
In the present work, we consider four different RMF parametrizations,
NL3~\cite{NL3}, TM1~\cite{TM1}, FSU~\cite{FSU}, and IUFSU~\cite{IUFSU},
so that we can examine the dependence of results on the RMF parametrization.
It is known that these RMF parametrizations are successful in reproducing
the ground-state properties of finite nuclei including unstable ones.
The NL3 model includes nonlinear terms of the $\sigma$ meson only,
while the TM1 model includes nonlinear terms for both $\sigma$
and $\omega$ mesons. An additional $\omega$-$\rho$ coupling term is added
in the FSU and IUFSU models; it plays an important role in
modifying the density dependence of the symmetry energy and
affecting the neutron-star properties~\cite{FSU,IUFSU,Horo01,Horo03,Mene11,Prov13}.
The IUFSU parametrization was developed from FSU by reducing the neutron-skin
thickness of $^{208}$Pb and increasing the maximum neutron-star mass in
the parameter fitting~\cite{IUFSU}.
The TM1 model was successfully used to construct the equation of
state for supernova simulations and neutron-star calculations~\cite{Shen02,Shen11}.
For completeness, we present the parameter sets and corresponding properties
of nuclear matter and finite nuclei for these RMF models in Table~\ref{tab:1}.
It is noticeable that the symmetry energy slope $L$ of the RMF models
ranges from a low value of $47.2$ MeV for IUFSU to a high value of $118.2$ MeV for NL3.
Therefore, it is possible to investigate the influence of $L$ on the results obtained
by using these different RMF models.

\subsection{Relativistic Thomas-Fermi approximation for hot nuclei}
\label{sec:2.2}

We use the relativistic Thomas-Fermi approximation to describe hot nuclei by 
employing the subtraction procedure as described in Refs.~\cite{BLV85,TF87}.
The subtraction procedure was proposed in order to isolate the hot nucleus
from the surrounding gas, so that the resulting properties of the nucleus
could be independent of the size of the box in which the calculation was performed.
At given temperature and chemical potentials, there exist two solutions
to the equations in the RMF model derived from the Lagrangian density~(\ref{eq:LRMF}),
one corresponding to the nuclear liquid phase with the surrounding
gas ($NG$) and the other corresponding to the nucleon gas alone ($G$).
The density profile of the nucleus ($N$) is then obtained by subtracting
the gas density from that of the liquid-plus-gas phase.
Without the inclusion of the Coulomb interaction in the Lagrangian
density~(\ref{eq:LRMF}), the gas phase is just diluted uniform nuclear
matter, while the liquid-plus-gas phase is an uncharged
nucleus in equilibrium with surrounding nucleon gas.
The inclusion of the Coulomb interaction leads to a difficulty in describing
hot nuclei. Because the Coulomb repulsion increases with the box size,
it drives protons to the border and finally results in a divergence.
To overcome this difficulty, the author of Ref.~\cite{BLV85} proposed 
calculating the Coulomb potential from the subtracted proton density,
which is the proton density of the isolated nucleus.
This implies that protons in the gas phase do not contribute to the Coulomb
potential, but they can be influenced by the Coulomb potential of the nucleus.
This prescription for the inclusion of the Coulomb interaction
is quite successful in describing hot nuclei, and, as a result,
properties of the hot nucleus are independent of the box size.

Using the relativistic Thomas-Fermi approximation with the subtraction procedure,
we study a hot nucleus based on the thermodynamic potential of the isolated nucleus,
which is defined by
\begin{equation}
\Omega=\Omega^{NG}-\Omega^{G}+E_{C},
\label{eq:ON}
\end{equation}
where $\Omega^{NG}$ and $\Omega^{G}$ are the nucleonic thermodynamic potentials
in the $NG$ and $G$ phases, respectively.
We employ the RMF model to calculate the thermodynamic potential
$\Omega^{a}$ ($a=NG$ or $G$), which can be written as
\begin{eqnarray}
\label{eq:Ormf}
\Omega^a=E^a-TS^a-\sum_{i=p,n}\mu_{i}N^a_{i}.
\end{eqnarray}
Here, the energy $E^a$, entropy $S^a$, and particle number $N^a_i$
in the phase $a$ are obtained by
\begin{eqnarray}
E^a &=&\int \varepsilon^a (r)d^3 r,
\label{eq:Ea}\\
S^a &=&\int s^a (r) d^3 r,
\label{eq:Sa} \\
N^a_{i} &=&\int n^a_{i}(r) d^3 r,
\label{eq:Nia}
\end{eqnarray}
where $\varepsilon^a(r)$,  $s^a(r)$, and $n^a_{i}(r)$ are the local energy
density, entropy density, and particle number density in the RMF model
given by Eqs.~(\ref{eq:ermf}), (\ref{eq:srmf}), and (\ref{eq:nirmf}), respectively.
The Coulomb energy is calculated from the subtracted proton density as
\begin{eqnarray}
\label{eq:EC}
E_{C}=\int \left[ e \left( n_{p}^{NG}-n_{p}^{G}\right) A_0
                  -\frac{1}{2}(\nabla A_0 )^{2}\right] d^{3}r,
\end{eqnarray}
where $A_0$ is the electrostatic potential.

The equilibrium state of the isolated nucleus can be obtained by
minimization of the thermodynamic potential $\Omega$ defined in Eq.~(\ref{eq:ON}).
The meson mean fields in the $NG$ phase satisfy the variational equation
\begin{eqnarray}
\label{eq:MNG}
\frac{\delta \Omega }{\delta \phi^{NG}}=0,
\hspace{1cm} \phi^{NG} =\sigma^{NG},\,\omega^{NG},\,\rho^{NG},
\end{eqnarray}
which leads to the following equations of motion for meson mean fields
in the $NG$ phase:
\begin{subequations}
\label{eq:png}
\begin{eqnarray}
&&-\nabla^{2}\sigma^{NG}+m_{\sigma}^{2}\sigma^{NG}
+g_{2}\left(\sigma^{NG}\right)^{2}+g_{3}\left(\sigma^{NG}\right)^{3}
=-g_{\sigma}\left(n_{s,p}^{NG}+n_{s,n}^{NG}\right), \\
&&-\nabla^{2}\omega^{NG}+m_{\omega}^{2}\omega^{NG}
+c_3\left(\omega^{NG}\right)^{3}
+2\Lambda_{\text{v}}g_{\omega}^{2}g_{\rho}^{2}\omega^{NG}\left(\rho^{NG}\right)^{2}
=g_{\omega}\left(n_{p}^{NG}+n_{n}^{NG}\right) , \\
&&-\nabla^{2}\rho^{NG}+m_{\rho}^{2}\rho^{NG}
+2\Lambda_{\text{v}}g_{\omega}^{2}g_{\rho}^{2}\left(\omega^{NG}\right)^{2}\rho^{NG}
=\frac{g_{\rho}}{2}\left( n_{p}^{NG}-n_{n}^{NG}\right) .
\end{eqnarray}
\end{subequations}
The occupation probability $f_{i+}^{k,NG}$ ($f_{i-}^{k,NG}$)
of species $i$ ($i=p$ or $n$) can be derived from the variational equation,
\begin{eqnarray}
\frac{\delta \Omega }{\delta f_{i\pm}^{k,NG}}=0,
\end{eqnarray}
which results in the Fermi-Dirac distribution of particles and antiparticles,
{\footnotesize
\begin{eqnarray}
f_{i\pm}^{k,NG} &=&\left\{1+\exp \left[ \left( \sqrt{k^{2}+
 \left( M^{\ast,NG}\right)^{2}}+g_{\omega }\omega^{NG}
 +\frac{g_{\rho }}{2}\tau_{3}\rho^{NG}+e\frac{\tau_{3}+1}{2}A_0
 \mp \mu_{i}\right) /T\right] \right\}^{-1}.
\end{eqnarray}
}
Similarly, we obtain the equations of motion for meson mean fields
in the $G$ phase as
\begin{subequations}
\label{eq:pg}
\begin{eqnarray}
&&-\nabla^{2}\sigma^{G}+m_{\sigma}^{2}\sigma^{G}
+g_{2}\left(\sigma^{G}\right) ^{2}+g_{3}\left(\sigma^{G}\right)^{3}
=-g_{\sigma }\left(n_{s,p}^{G}+n_{s,n}^{G}\right), \\
&&-\nabla^{2}\omega^{G}+m_{\omega}^{2}\omega^{G}
+c_3\left(\omega^{G}\right)^{3}
+2\Lambda_{\text{v}}g_{\omega}^{2}g_{\rho}^{2}\omega^{G}\left(\rho^{G}\right)^{2}
=g_{\omega }\left(n_{p}^{G}+n_{n}^{G}\right) , \\
&&-\nabla^{2}\rho^{G}+m_{\rho}^{2}\rho^{G}
+2\Lambda_{\text{v}}g_{\omega}^{2}g_{\rho}^{2}\left(\omega^{G}\right)^{2}\rho^{G}
=\frac{g_{\rho}}{2}\left( n_{p}^{G}-n_{n}^{G}\right),
\end{eqnarray}
\end{subequations}
and the occupation probability in the $G$ phase as
{\small
\begin{eqnarray}
f_{i\pm}^{k,G} &=&\left\{1+\exp \left[ \left( \sqrt{k^{2}+
 \left( M^{\ast,G}\right)^{2}}+g_{\omega}\omega^{G}
 +\frac{g_{\rho}}{2}\tau_{3}\rho^{G}+e\frac{\tau_{3}+1}{2}A_0
 \mp \mu_{i}\right) /T\right] \right\}^{-1}.
\end{eqnarray}
}
In the equations for meson mean fields, $n_{s,i}^{a}$ and $n_{i}^{a}$
denote, respectively, the scalar and number densities of species $i$ ($i=p$ or $n$)
in the $a$ ($a=NG$ or $G$) phase.
The number density $n_{i}^{a}$ is calculated from Eq.~(\ref{eq:nirmf}),
while the scalar density $n_{s,i}^{a}$ is given by
\begin{equation}
\label{eq:nsrmf}
 n_{s,i}^{a}(r)=\frac{1}{\pi^2}
       \int_0^{\infty} dk\,k^2\,\frac{M^{\ast,a}}{\sqrt{k^{2}+\left( M^{\ast,a}\right)^{2}}}
       \left(f_{i+}^{k,a}+f_{i-}^{k,a}\right).
\end{equation}
By minimizing $\Omega$ with respect to the electrostatic potential $A_0$,
we obtain the Poisson equation for  $A_0$ as
\begin{eqnarray}
\label{eq:EQA}
-\nabla^{2}A_0=e\left( n_{p}^{NG}-n_{p}^{G}\right) .
\end{eqnarray}
The inclusion of the Coulomb energy in $\Omega$ leads to a coupling
between the two sets of equations for the $NG$ and $G$ phases.
Therefore, the coupled equations (\ref{eq:png}), (\ref{eq:pg}),
and~(\ref{eq:EQA}) should be solved simultaneously at given
temperature $T$ and chemical potentials $\mu_p$ and $\mu_n$.
To solve Eqs. (\ref{eq:png}) and (\ref{eq:pg}),
we take the boundary conditions for meson mean fields
in phase $a$ ($a=NG$ or $G$) as
\begin{eqnarray}
\label{eq:BCM}
\frac{d \phi^{a} }{dr} (r=0) = 0, \hspace{1cm}
\frac{d \phi^{a} }{dr} (r=R) = 0, \hspace{1cm}
\hspace{1cm} \phi^{a} =\sigma^{a},\,\omega^{a},\,\rho^{a},
\end{eqnarray}
where $r=0$ and $r=R$ represent, respectively, the center and
the edge of a spherical box with radius $R$.
For the electrostatic potential $A_0$, the boundary conditions
are taken as
\begin{eqnarray}
\label{eq:BCA}
\frac{d A_0 }{dr} (r=0) = 0, \hspace{1cm}
A_0 (r=R) = \frac{e N_p}{4\pi R},
\end{eqnarray}
where $N_p$ is the proton number of the isolated nucleus given by
Eq.~(\ref{eq:Ni}) below, and $e=\sqrt{4\pi/137}$ is
the electromagnetic coupling constant.
The box radius $R$ is generally taken to be sufficiently large
(about 15--20 fm) so that the results of the isolated nucleus
could be independent of the box size.
In the present calculations, we set $R=20$ fm, and we have
checked that the resulting properties of the nucleus remain
unchanged when varying $R$ from $15$ to $20$ fm for $T \leq 8$ MeV.

For a nucleus with $N_p$ protons and $N_n$ neutrons at temperature $T$,
the proton and neutron chemical potentials $\mu_p$ and $\mu_n$ can be
determined from given $N_p$ and $N_n$.
Once the chemical potentials are known, the occupation probabilities
and density distributions can be obtained easily.
In practice, we solve self-consistently the coupled
set of Eqs. (\ref{eq:png}), (\ref{eq:pg}), and~(\ref{eq:EQA})
under the constraints of given $N_p$ and $N_n$.
After getting the solutions for
the $NG$ and $G$ phases, we can extract the properties of hot nuclei
based on the subtraction procedure.
The proton and neutron numbers, $N_p$ and $N_n$, are given by
\begin{eqnarray}
N_{i} = N^{NG}_i - N^{G}_i=\int n_i(r) d^3 r,
\hspace{1cm} i=p,\, n,
\label{eq:Ni}
\end{eqnarray}
where $n_{i}(r)=n^{NG}_i(r)-n^{G}_i(r)$ is the local density
of the isolated nucleus, which decreases to zero at large distances.
Therefore, physical quantities of the isolated nucleus could be
independent of the size of the box in which the calculation is performed.
The total energy of the hot nucleus is given by
\begin{eqnarray}
\label{eq:EN}
E=E^{NG}-E^{G}+E_{C},
\end{eqnarray}
where $E^{NG}$ and $E^{G}$ are the nucleonic energies without the Coulomb
interaction in the $NG$ and $G$ phases, which are calculated
from Eq.~(\ref{eq:Ea}). The Coulomb energy $E_{C}$ is given by
Eq.~(\ref{eq:EC}). For a nucleus at temperature $T$,
its excitation energy is defined as
\begin{eqnarray}
\label{eq:Estar}
E^{\ast}(T)=E(T)-E(T=0).
\end{eqnarray}
The entropy and other extensive quantities of the isolated nucleus
can be calculated by subtracting the contribution of the $G$ phase
from that of the $NG$ phase.

\subsection{Symmetry energy of finite nuclei}
\label{sec:2.3}

The symmetry energy is a key quantity in the study of exotic nuclei,
heavy-ion collisions, and astrophysical phenomena~\cite{PR05,PR07,LiBA08}.
For infinite nuclear matter, the symmetry energy is defined by
expanding the energy per particle, $\epsilon(n,\alpha)$, in terms of the isospin
asymmetry parameter, $\alpha=(n_n-n_p)/n$, as
\begin{eqnarray}
\label{eq:ESNM}
\epsilon(n,\alpha)=\epsilon(n,0)+a^v_{\rm{sym}}(n)\alpha^2+O(\alpha^4),
\end{eqnarray}
where $n=n_n+n_p$ is the nucleon number density.
$a^v_{\rm{sym}}(n)$ is the symmetry energy coefficient of nuclear matter at density $n$,
and its value at saturation density $n_0$ is about $30$--$34$
MeV~\cite{PLB12,Sub14b,LDM06}.
On the other hand, the symmetry energy of finite nuclei
can be defined based on the Bethe-Weizs\"{a}cker mass formula,
which gives the expression for the binding energy per particle as
\begin{eqnarray}
\label{eq:ESFN}
\frac{E}{A}&=&\epsilon(A,Z)
  =\epsilon_{\rm{vol}} + \epsilon_{\rm{surf}} + \epsilon_{\rm{sym}}
         + \epsilon_{\rm{Coul}} + \epsilon_{\rm{pair}} \notag\\
 &=&-a_{\rm{vol}} + a_{\rm{surf}} A^{-1/3} + a_{\rm{sym}}\frac{(N-Z)^2}{A^2}
    +a_{\rm{Coul}}\frac{Z^2}{A^{4/3}}+\epsilon_{\rm{pair}},
\end{eqnarray}
where $N$ and $Z$ are the neutron and proton numbers, respectively,
while $A=N+Z$ is the mass number.
The volume, surface, symmetry, and Coulomb energies per particle are
denoted by $\epsilon_{\rm{vol}},\, \epsilon_{\rm{surf}},\,
\epsilon_{\rm{sym}},\,$ and $\epsilon_{\rm{Coul}}$, respectively, while
the last term $\epsilon_{\rm{pair}}$ represents the pairing correction.
Here, $a_{\rm{sym}}$ is the symmetry energy coefficient of finite nuclei,
which is generally dependent on the mass number $A$.
The symmetry energy of finite nuclei includes the volume and surface terms,
and they are related by~\cite{SEA03}
\begin{eqnarray}
\label{eq:avas1}
a_{\rm{sym}} (A)=\frac{a^v_{\rm{sym}}}
 {1+\left(a^v_{\rm{sym}}/a^s_{\rm{sym}}\right) A^{-1/3}},
\end{eqnarray}
where $a^v_{\rm{sym}}$ and $a^s_{\rm{sym}}$ are the volume
and surface symmetry energy coefficients, respectively.
In the limit of large $A$, $a_{\rm{sym}} (A)$
can be expanded in power of $A^{-1/3}$, which results in the relation~\cite{LDM06}
\begin{eqnarray}
\label{eq:avas2}
a_{\rm{sym}} (A)=a^v_{\rm{sym}}+a_{\rm{ssym}} A^{-1/3},
\end{eqnarray}
with $a_{\rm{ssym}} = -(a^v_{\rm{sym}})^2/ (a^s_{\rm{sym}})$.
The volume symmetry energy coefficient $a^v_{\rm{sym}}$ is identified as
the symmetry energy coefficient of infinite nuclear matter at saturation
density, and its value is around $30$--$34$ MeV as mentioned above.
The ratio of volume and surface coefficients, $a^v_{\rm{sym}}/a^s_{\rm{sym}}$,
was found to be in a range of $\sim$0.4--$3.4$ for Skyrme interactions,
as shown in Fig.~15 of Ref.~\cite{SEA09}.
In the present work, we calculate the symmetry energy coefficient of a nucleus
with mass number $A$ at temperature $T$ from a different method~\cite{SEA02,Sub12}:
\begin{equation}
\label{eq:EAFN}
a_{\rm{sym}}(A,T) = \left[\epsilon_b(A,X_{1},T)-\epsilon_b(A,X_{2},T)\right]/
                    \left(X_{1}^2-X_{2}^2\right),
\end{equation}
where $X_{1}$ and $X_{2}$ are the neutron excesses of a pair of
nuclei having the same mass number $A$ but different proton number $Z$.
For a nucleus with $Z$ protons and $N$ neutrons,
the neutron excess is defined by $X=(N-Z)/A$.
Here, $\epsilon_b=\epsilon-\epsilon_C$ is the energy per particle
obtained by subtracting the Coulomb part.
The total energy per particle, $\epsilon=E/A$, is calculated
from Eq.~(\ref{eq:EN}), while the Coulomb energy per particle,
$\epsilon_C=E_C/A$, is obtained from Eq.~(\ref{eq:EC}).
In practice, we choose a pair of even-even nuclei with the same $A$
but different $Z$ for the calculation of
$a_{\rm{sym}}(A,T)$ used in Eq.~(\ref{eq:EAFN}).
According to the definition given by Eq.~(\ref{eq:EAFN}),
the resulting $a_{\rm{sym}}(A,T)$ would also depend on
the choice of nuclear pair, namely, the choice of $Z_{1}$
and $Z_{2}$. This dependence has been discussed in Ref.~\cite{Sub12}
by using the nonrelativistic Thomas-Fermi approximation with Skyrme
effective interactions.
In the present study, we will employ the relativistic Thomas-Fermi
approximation with the RMF model for nuclear interactions
to extract the symmetry energy coefficient of finite nuclei from Eq.~(\ref{eq:EAFN}).
We will also evaluate the symmetry energy coefficient from uncharged nuclei;
namely, the Coulomb interaction is switched off, so that the symmetry energy
coefficient extracted would not be hindered by the Coulomb interaction.
In the next section, we will present and discuss our results
and make a comparison with those obtained in the nonrelativistic approach.

%%%%%%%%%%%%%%%%%%%%%%%%%%%%%%%%%%%%%%%%%%%%%%%%%%%%%%%%%%%%%%%%%%%%%
\section{Results and discussion}
\label{sec:3}

\begin{figure}[htb]
\includegraphics[bb=15 60 540 680, width=8.6 cm,clip]{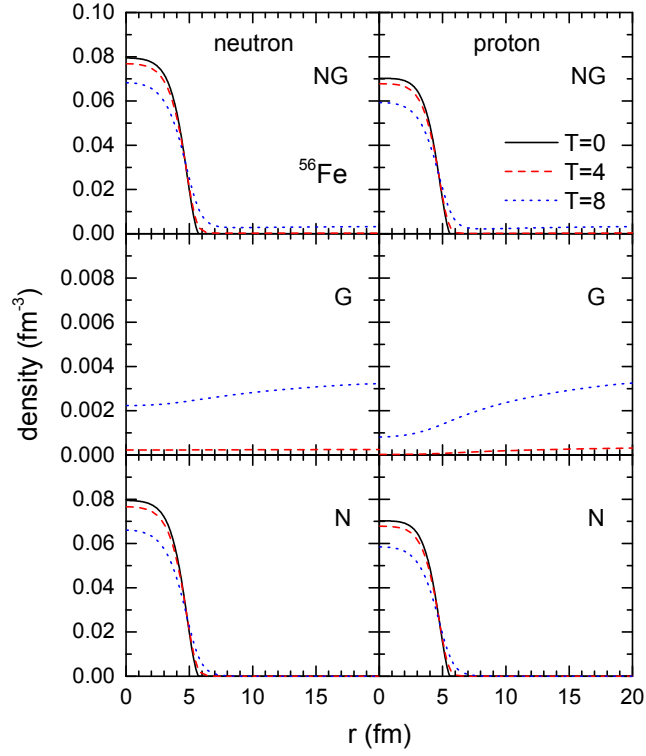}
\caption{(Color online) Density distributions of neutrons (left panels)
and protons (right panels) for $^{56}$Fe at $T=0$, 4, and 8 MeV obtained
using the TM1 parametrization. The results of the nuclear liquid-plus-gas
phase ($NG$), nucleon gas phase ($G$), and subtracted nuclear liquid
phase ($N$) are shown in the top, middle, and bottom panels, respectively.}
\label{fig:1dis56}
\end{figure}

\begin{figure}[htb]
\includegraphics[bb=15 60 540 680, width=8.6 cm,clip]{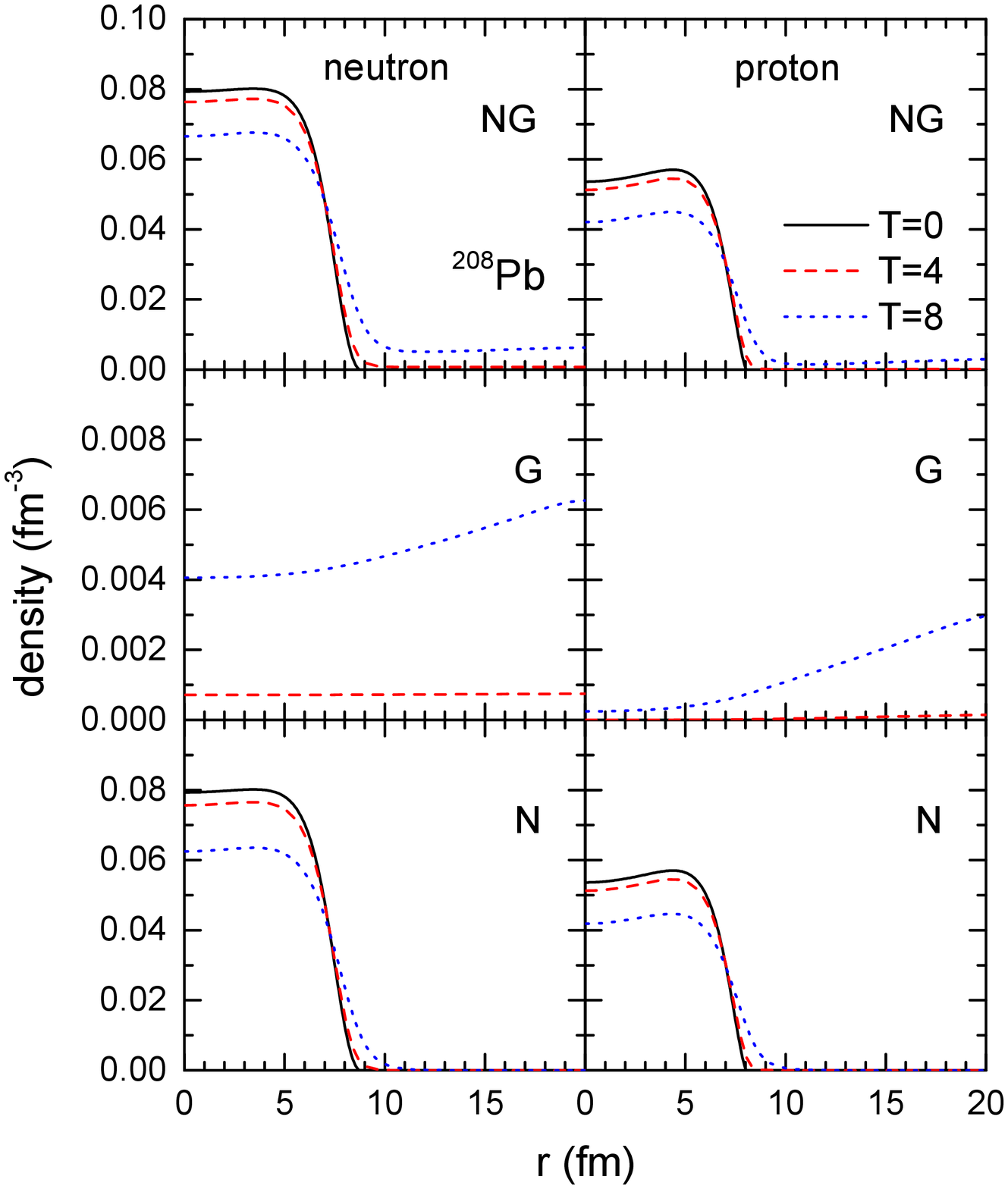}
\caption{(Color online) Same as Fig.~\ref{fig:1dis56}, but for $^{208}$Pb.}
\label{fig:2dis208}
\end{figure}

In this section, we investigate properties of hot nuclei within the
relativistic Thomas-Fermi approximation by employing the subtraction procedure.
For a nucleus with $N_p$ protons and $N_n$ neutrons at temperature $T$,
we solve the coupled set of Eqs. (\ref{eq:png}), (\ref{eq:pg}), and~(\ref{eq:EQA})
with the chemical potentials $\mu_p$ and $\mu_n$ constrained by given $N_p$
and $N_n$. After getting the solutions for the $NG$ and $G$ phases, we can obtain
physical properties of the isolated nucleus ($N$) using the
subtraction procedure. In Figs.~\ref{fig:1dis56} and~\ref{fig:2dis208},
we display the density distributions of neutrons and protons for $^{56}$Fe
and $^{208}$Pb at $T=0$, 4, and 8 MeV obtained by using the TM1 parametrization.
From top to bottom, we show, respectively, the results of the nuclear liquid-plus-gas
phase ($NG$), nucleon gas phase ($G$), and subtracted nuclear liquid phase ($N$).
It is clear that the densities of the isolated nucleus ($N$)
vanish at large distances, and, as a result, physical quantities of the
nucleus could be independent of the box size.
At zero temperature, the proton and neutron densities of the $G$ phase
are found to be exactly zero, while their values are finite but very small
at finite temperature. As temperature increases, the neutron and proton
densities of the $G$ phase increase significantly. Moreover, densities
at the center of the nucleus are reduced and the nuclear surface becomes more diffuse
with increasing $T$. As can be seen in the middle panels of Figs.~\ref{fig:1dis56}
and~\ref{fig:2dis208}, the densities of the $G$ phase are obviously
polarized due to the inclusion of the Coulomb interaction.
The nucleon distributions at the center of a heavy nucleus
such as $^{208}$Pb are significantly affected by the repulsive Coulomb potential;
namely, the nucleon densities (especially the proton densities)
at the center of the nucleus are slightly lower than those
at the surface region, as shown in the bottom panel of Fig.~\ref{fig:2dis208}.
These tendencies are consistent with those obtained in nonrelativistic
approaches~\cite{BLV85,TF87}.

\begin{figure}[htb]
\includegraphics[bb=30 50 550 750, width=8.6 cm,clip]{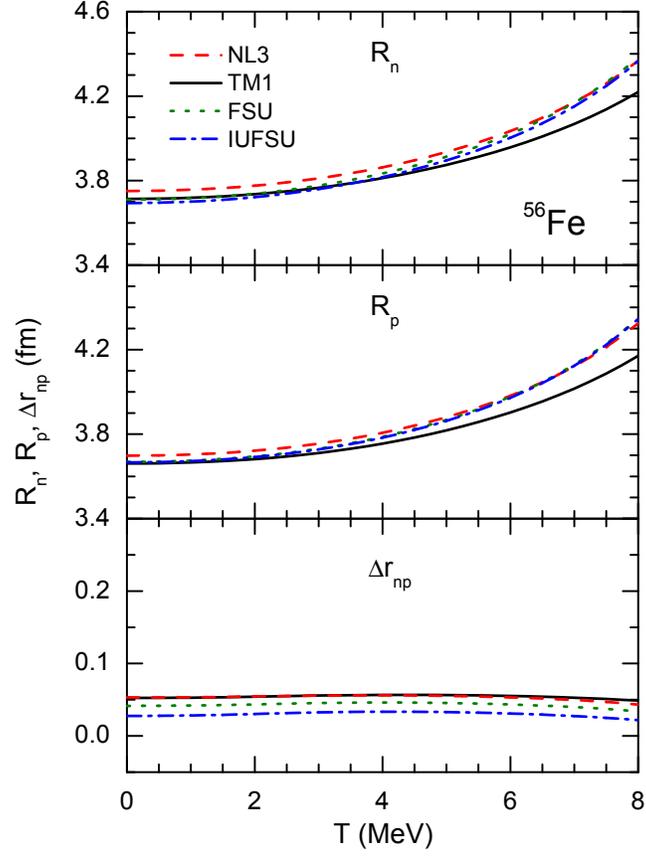}
\caption{(Color online) The rms radii of neutrons and protons,
$R_n$ and $R_p$, and the neutron-skin thickness, $\Delta r_{np}=R_n-R_p$,
as a function of the temperature $T$ for $^{56}$Fe obtained using
four different RMF parametrizations.}
\label{fig:3r56}
\end{figure}

\begin{figure}[htb]
\includegraphics[bb=30 50 550 750, width=8.6 cm,clip]{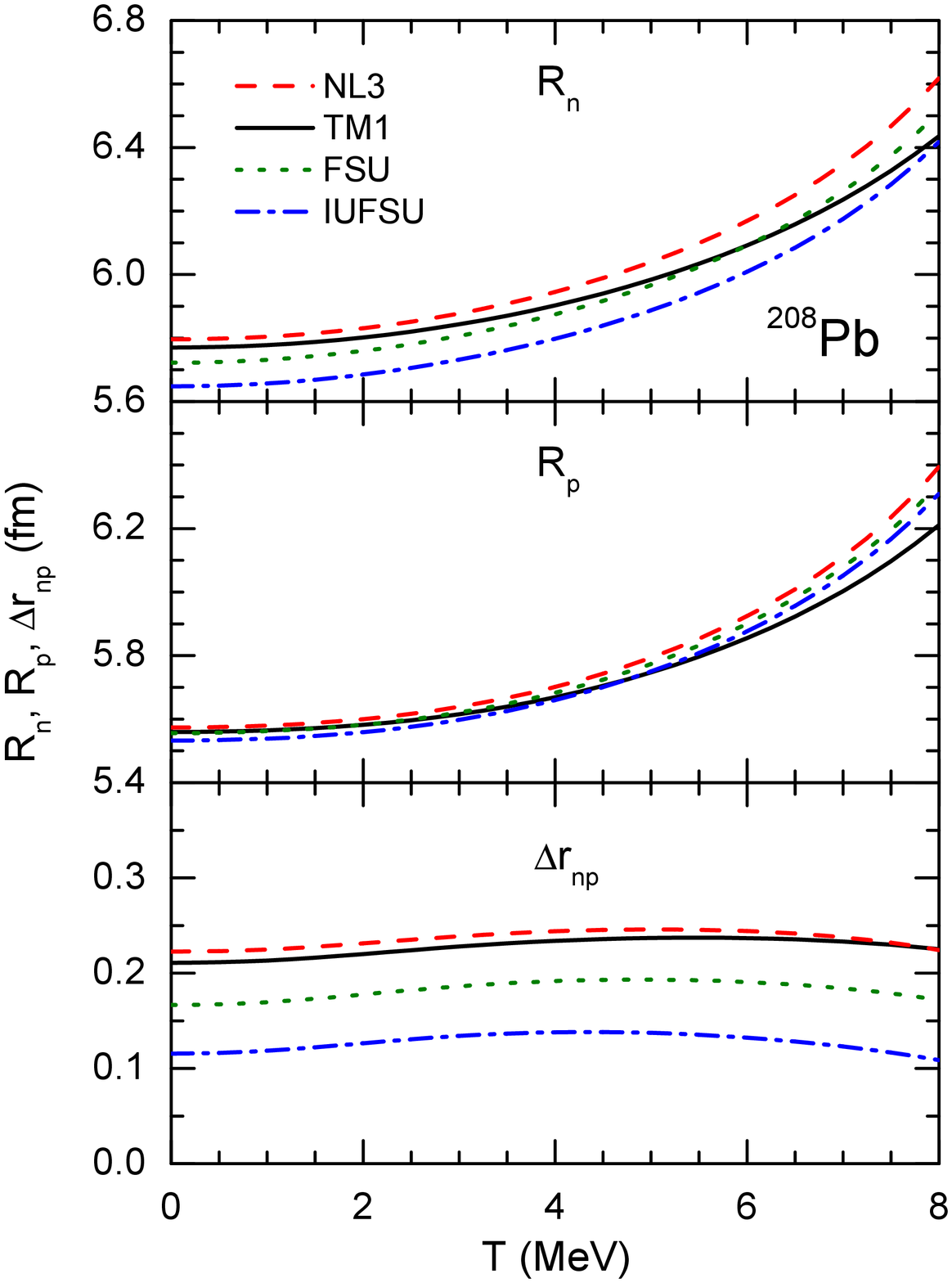}
\caption{(Color online) Same as Fig.~\ref{fig:3r56}, but for $^{208}$Pb.}
\label{fig:4r208}
\end{figure}

We show in Figs.~\ref{fig:3r56} and~\ref{fig:4r208} the root-mean-square (rms)
radii of neutrons and protons, $R_n$ and $R_p$, and their difference known as
the neutron-skin thickness, $\Delta r_{np}=R_n-R_p$, as a function of the
temperature $T$ for $^{56}$Fe and $^{208}$Pb, respectively.
It is well known that there is a correlation between the neutron-skin
thickness $\Delta r_{np}$ and the symmetry energy slope $L$~\cite{FSU,IUFSU,Horo01,Horo03}.
In order to study the temperature dependence of this correlation,
we calculate $\Delta r_{np}$ by using four different RMF parametrizations,
namely, NL3, TM1, FSU, and IUFSU, which cover a wide range of $L$ as
listed in Table~\ref{tab:1}.
It is shown that a larger $L$ favors a larger $\Delta r_{np}$,
which does not change much with increasing $T$.
The values of $\Delta r_{np}$ for $^{208}$Pb at $T=0$ are presented in Table~\ref{tab:1}.
On the other hand, both $R_n$ and $R_p$ are found to increase significantly
with increasing $T$. This is because the nucleon distributions become more
diffuse at higher temperature, as shown in Figs.~\ref{fig:1dis56} and~\ref{fig:2dis208}.
The increase of $R_n$ and $R_p$ in TM1 is somewhat slower than that in
the other three cases, which can be roughly explained by a relatively small
decrease in the saturation density of nuclear matter. With increasing $T$,
the saturation density of nuclear matter obtained in TM1
deceases more slowly compared to that obtained in other three cases,
and, as a result, the size of a nuclear drop with the same mass number $A$
would increase more slowly with temperature, as shown by black-solid lines
in Figs.~\ref{fig:3r56} and~\ref{fig:4r208}.
The temperature dependence of $R_n$ and $R_p$ obtained in this work is
comparable to that shown in Ref.~\cite{TF87}.
For the influence of $L$ on $R_n$ and $R_p$, it is shown that $R_n$ of heavy nuclei is
more sensitive to $L$ than $R_p$, which is related to the correlation
between $\Delta r_{np}$ and $L$.

\begin{figure}[htb]
\includegraphics[bb=35 50 510 410, width=8.6 cm,clip]{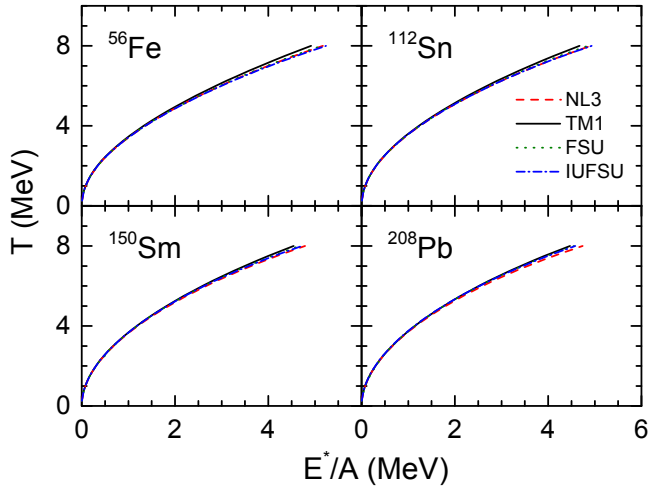}
\caption{(Color online) Temperature $T$ as a function of the
excitation energy per particle, $E^{\ast}/A$ (caloric curve),
for $^{56}$Fe, $^{112}$Sn, $^{150}$Sm, and $^{208}$Pb
obtained using four different RMF parametrizations.}
\label{fig:5TE}
\end{figure}

The excitation energies of hot nuclei are calculated from Eq.~(\ref{eq:Estar})
using the four different RMF parametrizations listed in Table~\ref{tab:1}.
We plot in Fig.~\ref{fig:5TE} the temperature $T$ as a function of
the excitation energy per particle, $E^{\ast}/A$ (the so-called caloric curve),
for four representative nuclei, $^{56}$Fe, $^{112}$Sn, $^{150}$Sm, and $^{208}$Pb.
It is shown that different RMF parametrizations produce very similar results in all cases.
This is because all of these models have been fitted to experimental masses
of finite nuclei covering a wide range of atomic numbers.
Therefore, binding energies obtained by these models are close to
each other. As one can see in Fig.~\ref{fig:5TE}, $E^{\ast}/A$ increases
slowly at low temperature, while it rises more rapidly as $T$ increases.
It is known that there exists a limiting temperature $T_{\rm{lim}}$
for a hot nucleus, above which the nucleus becomes unstable due to the Coulomb
interaction~\cite{BLV85,Tlim85,TF87,Tlim89}.
The limiting temperature $T_{\rm{lim}}$ depends on the nucleus,
and it generally decreases with increasing mass number $A$ and with
increasing charge-to-mass ratio $Z/A$~\cite{TF87}.
It was found in the finite-temperature Hartree-Fock calculation~\cite{BLV85}
and the nonrelativistic Thomas-Fermi approximation~\cite{TF87}
that $T_{\rm{lim}}$ is about 8--10 MeV depending on the nucleus
and effective interactions. In the present study, we find that
the instability of hot nuclei caused by the Coulomb
interaction occurs at $T>8$ MeV. Therefore, we perform our calculation for hot
nuclei up to $T \sim 8$ MeV.

\begin{figure}[htb]
\includegraphics[bb=25 25 515 595, width=8.6 cm,clip]{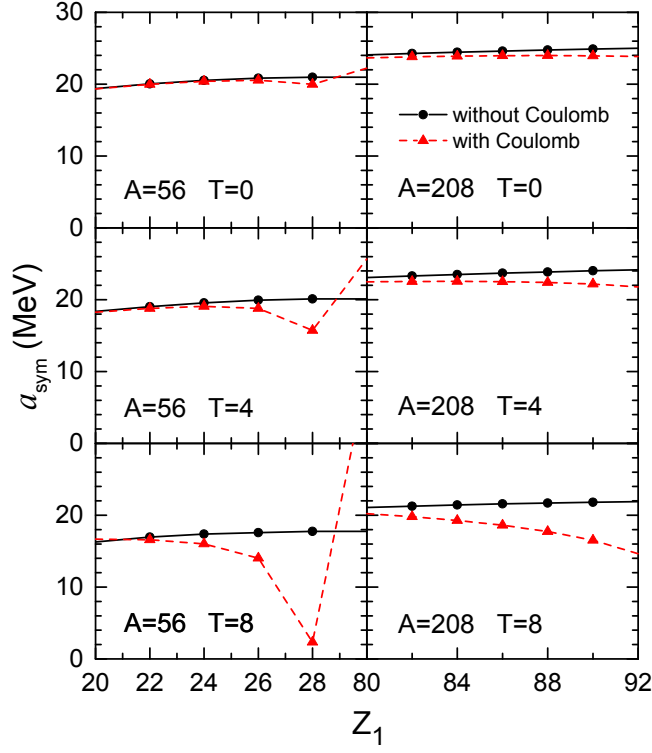}
\caption{(Color online) Symmetry energy coefficient $a_{\rm{sym}}$
as a function of $Z_1$ for $A=56$ and $A=208$ at $T=0$, $4$, and $8$ MeV
obtained using the TM1 parametrization. The nuclear pair used for
calculating $a_{\rm{sym}}$ is chosen to be ($A$, $Z_1$) and
($A$, $Z_1-2$). The results extracted from uncharged
nuclei (without Coulomb) are compared with those calculated from
charged nuclei by subtracting the Coulomb energy (with Coulomb).}
\label{fig:6EAZ1TM1}
\end{figure}

\begin{figure}[htb]
\includegraphics[bb=25 25 515 595, width=8.6 cm,clip]{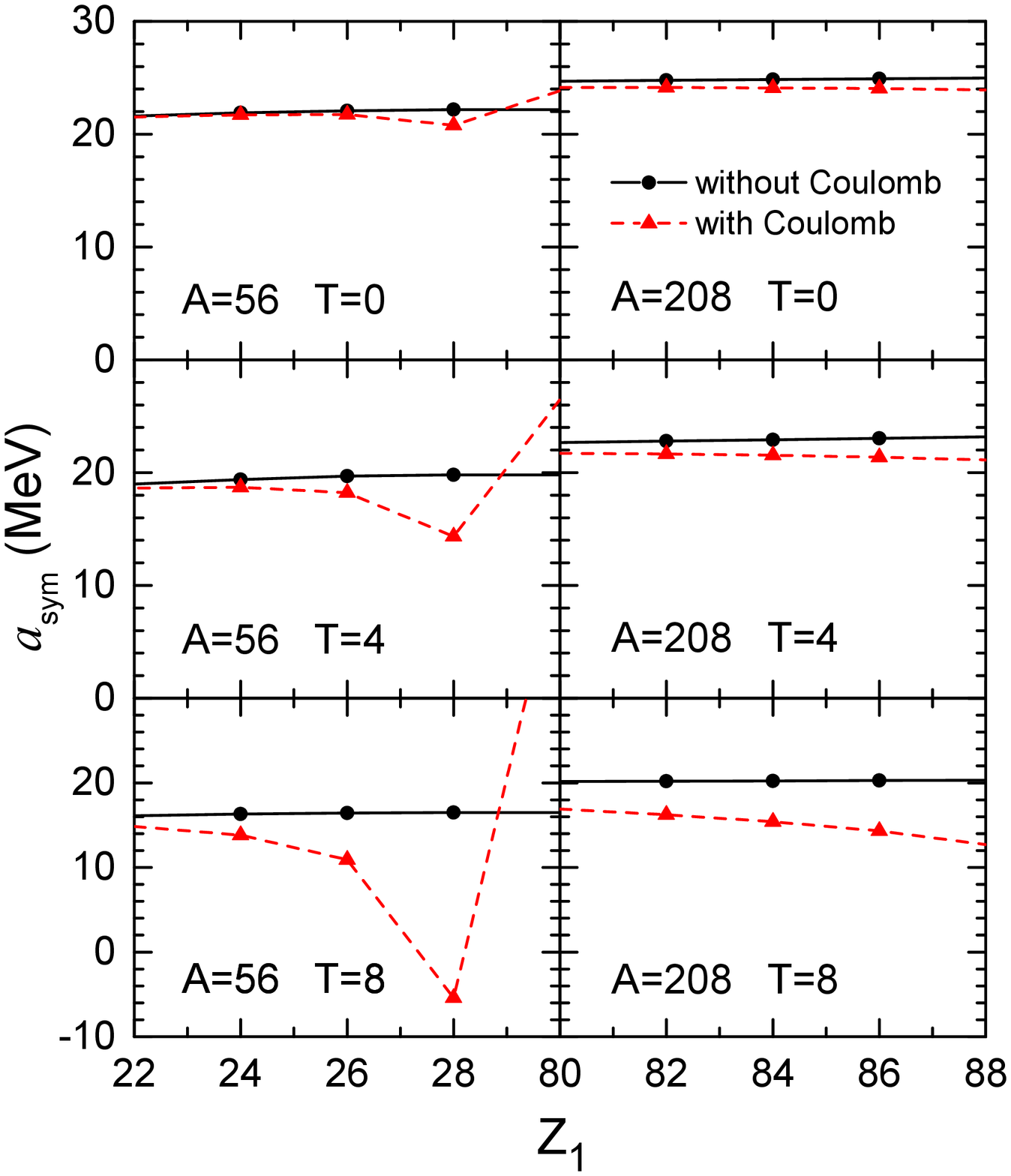}
\caption{(Color online) Same as Fig.~\ref{fig:6EAZ1TM1},
but for the FSU parametrization.}
\label{fig:7EAZ1FSU}
\end{figure}

\begin{figure}[htb]
\includegraphics[bb=30 30 510 435, width=8.6 cm,clip]{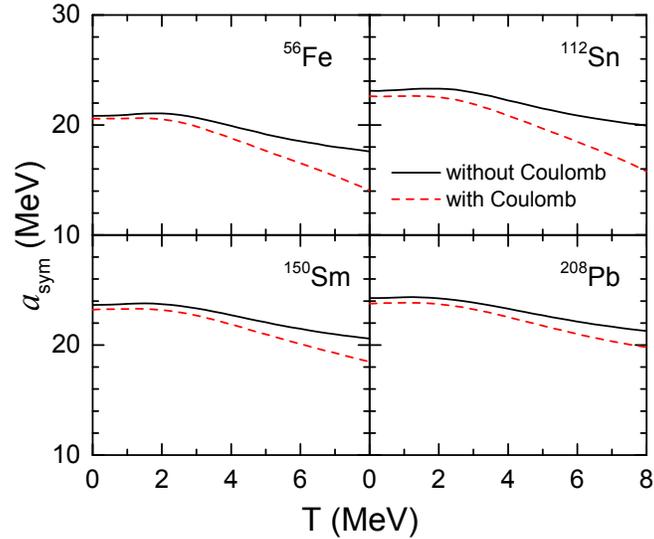}
\caption{(Color online) Temperature dependence of $a_{\rm{sym}}$
for $^{56}$Fe, $^{112}$Sn, $^{150}$Sm, and $^{208}$Pb
obtained using the TM1 parametrization.
The results obtained from uncharged nuclei (without Coulomb) are
compared with those calculated from charged nuclei by subtracting the
Coulomb energy (with Coulomb).}
\label{fig:8EATTM1}
\end{figure}

\begin{figure}[htb]
\includegraphics[bb=30 30 510 435, width=8.6 cm,clip]{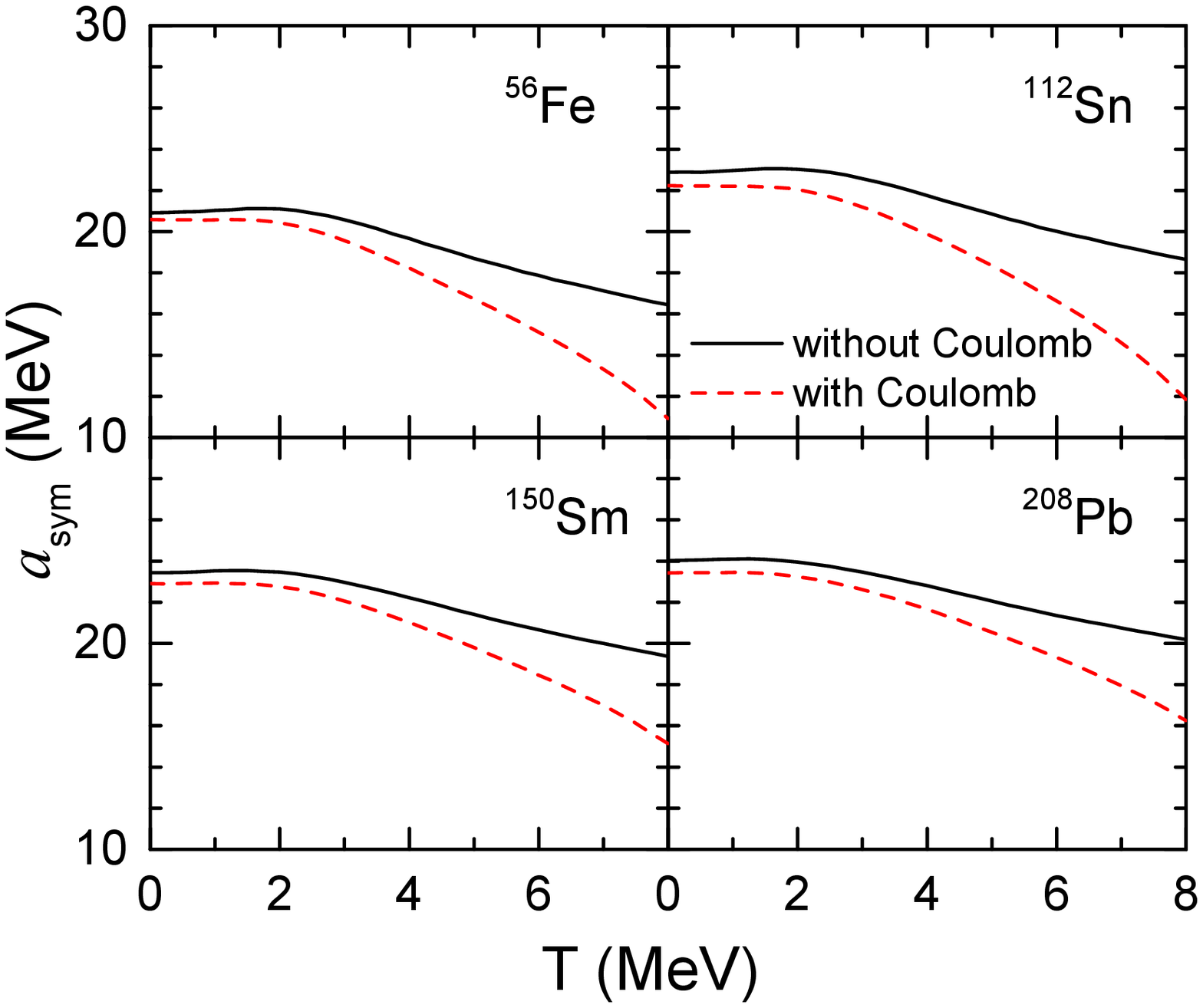}
\caption{(Color online) Same as Fig.~\ref{fig:8EATTM1},
but for the FSU parametrization.}
\label{fig:9EATFSU}
\end{figure}

We investigate the temperature dependence of the symmetry energy of finite
nuclei within the relativistic Thomas-Fermi approximation.
The symmetry energy coefficient of a nucleus with mass number $A$ at temperature $T$
is calculated from Eq.~(\ref{eq:EAFN}) by a pair of
nuclei with proton numbers $Z_1$ and $Z_2$. In practice, we choose the nuclear
pair ($A$, $Z_1$) and ($A$, $Z_2$) near the $\beta$-stability line
with $Z_1-Z_2=2$ for evaluating the symmetry energy coefficient $a_{\rm{sym}}$.
According to the liquid-drop model, the symmetry energy coefficient of finite nuclei
is dependent on the mass number $A$ due to the combination of volume
and surface contributions~\cite{SEA03,LDM06}, but it would not be very sensitive to the
choice of nuclear pair, namely, the choice of $Z_{1}$ and $Z_{2}$.
However, calculations done by De and Samaddar~\cite{Sub12}
show that $a_{\rm{sym}}$ depends sensitively on the choice of the 
nuclear pair (see Figs. 6, 7, and 8 of Ref.~\cite{Sub12}).
In order to test the dependence of $a_{\rm{sym}}$
on the choice of $Z_1$ with $Z_2=Z_1-2$, we plot in Figs.~\ref{fig:6EAZ1TM1}
and~\ref{fig:7EAZ1FSU} the coefficient $a_{\rm{sym}}$ by red-dashed lines
as a function of $Z_1$ for $A=56$ and $A=208$ at $T=0$, $4$, and $8$ MeV using
the TM1 and FSU parametrizations, respectively.
It is shown that the dependence of $a_{\rm{sym}}$ on $Z_1$
becomes more pronounced for higher temperature and close to isospin
symmetry. For the case of $T=8$ MeV and $A=56$, there is a sharp drop
in $a_{\rm{sym}}$ at $Z_1=28$, which is calculated from
the nuclear pair ($56$, $28$) and ($56$, $26$).
This value even becomes negative for the FSU parametrization, as shown
in the bottom-left panel of Fig.~\ref{fig:7EAZ1FSU}.
Negative $a_{\rm{sym}}$ extracted from the same nuclear pair
at $T=8$ MeV has also been presented in Fig.~6 of Ref.~\cite{Sub12}.
In principle, the symmetry energy coefficient could be extracted from
uncharged nuclei in which the Coulomb interaction is switched off~\cite{LDM06}.
This is because the symmetry energy is solely determined
by nuclear forces. It would be more appropriate to switch off the
Coulomb interaction in extracting the symmetry energy coefficient
of finite nuclei, rather than to subtract
the Coulomb energy after calculating charged nuclei as defined
by Eq.~(\ref{eq:EAFN}). This is because the Coulomb interaction can
significantly polarize nucleon densities and affect nuclear surface
properties~\cite{SEA03,Dong13}. As a result, the symmetry energy coefficient
calculated from Eq.~(\ref{eq:EAFN}) is hindered by the Coulomb
polarization. In Figs.~\ref{fig:6EAZ1TM1} and~\ref{fig:7EAZ1FSU},
we compare the results of $a_{\rm{sym}}$ extracted from uncharged
nuclei (without Coulomb) with those obtained from charged
nuclei (with Coulomb). As one can see, $a_{\rm{sym}}$ without the
Coulomb interaction becomes much less sensitive to the choice of nuclear pair
in comparison to that with the Coulomb interaction. We note that a lower sensitivity
to the nuclear pair is expected according to the liquid-drop model.
In Figs.~\ref{fig:8EATTM1} and~\ref{fig:9EATFSU}, we display and compare
the temperature dependence of $a_{\rm{sym}}$ obtained from charged
nuclei (with Coulomb) and uncharged nuclei (without Coulomb)
using the TM1 and FSU parametrizations, respectively.
The calculations are performed for four pairs of nuclei with $Z_2=Z_1-2$,
while the representative nuclei ($A$, $Z_1$) are chosen to be $^{56}$Fe,
$^{112}$Sn, $^{150}$Sm, and $^{208}$Pb.
It is shown that $a_{\rm{sym}}$ with the Coulomb interaction decreases more
rapidly than that without the Coulomb interaction, which may be related to a more
pronounced Coulomb polarization effect at higher temperature.
It has been pointed out in Ref.~\cite{Sub12} that a fast drop of
$a_{\rm{sym}}$ even results in negative values of $a_{\rm{sym}}$,
which is likely to arise from the Coulomb polarization of
nucleon densities. In the present study, we find that $a_{\rm{sym}}$
obtained from uncharged nuclei is less sensitive to the choice of
nuclear pair as expected and shows a smaller decrease than that
with the Coulomb interaction. The relatively weak temperature dependence of $a_{\rm{sym}}$
obtained without the Coulomb interaction seems to be more consistent with the behavior
of infinite nuclear matter~\cite{SEM07}.

\begin{figure}[htb]
\includegraphics[bb=20 30 510 435, width=8.6 cm,clip]{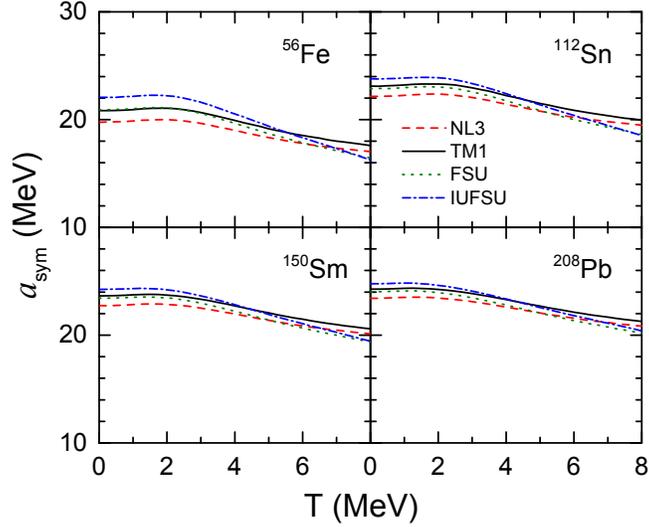}
\caption{(Color online) Temperature dependence of $a_{\rm{sym}}$
calculated from uncharged nuclei for $^{56}$Fe, $^{112}$Sn,
$^{150}$Sm, and $^{208}$Pb with four different RMF parametrizations.}
\label{fig:10EAall}
\end{figure}

In order to examine the impact of the RMF parametrization on the
results obtained, we calculate $a_{\rm{sym}}$ from uncharged nuclei
by using four different RMF parametrizations, NL3, TM1, FSU, and IUFSU.
We show in Fig.~\ref{fig:10EAall} the results calculated from four pairs
of nuclei with $Z_2=Z_1-2$, while the representative nuclei ($A$, $Z_1$)
are again chosen to be $^{56}$Fe, $^{112}$Sn, $^{150}$Sm, and $^{208}$Pb.
One can see that there are no large differences among these RMF
parametrizations, and all of them show a decreasing $a_{\rm{sym}}$
with increasing temperature.
It is shown that the largest $L$ of NL3 is associated with the
smallest $a_{\rm{sym}}$ at $T=0$ for all nuclei considered.
This is because the surface term for the model with a larger $L$ can
contribute more to the symmetry energy; namely,
the ratio of volume and surface coefficients, $a^v_{\rm{sym}}/a^s_{\rm{sym}}$,
appearing in Eq.~(\ref{eq:avas1}) has a bigger value for a larger $L$.
The correlation between $a^v_{\rm{sym}}/a^s_{\rm{sym}}$ and $L$
calculated from Skyrme interactions has been shown in Fig.~15
of Ref.~\cite{SEA09}. Here, we can calculate the ratio
$a^v_{\rm{sym}}/a^s_{\rm{sym}}$ at $T=0$ for each RMF parametrization
from Eq.~(\ref{eq:avas1}) using the results of $^{208}$Pb.
With $a^v_{\rm{sym}}$ given in Table~\ref{tab:1}
and $a_{\rm{sym}} (A=208, T=0)$ calculated from
the pair of uncharged nuclei ($208$, $82$) and ($208$, $80$),
we obtain that the ratio $a^v_{\rm{sym}}/a^s_{\rm{sym}}$ ranges
from $1.56$ for IUFSU to $3.54$ for NL3 (see Table~\ref{tab:1}).
The correlation between $a^v_{\rm{sym}}/a^s_{\rm{sym}}$ and $L$
obtained in the present study
is consistent with that shown in Fig.~15 of Ref.~\cite{SEA09}.
Due to the large value of $a^v_{\rm{sym}}/a^s_{\rm{sym}}$ for NL3,
we obtain a relatively small $a_{\rm{sym}} (A=208, T=0)=23.4$ MeV
for NL3 (see the bottom-right panel of Fig.~\ref{fig:10EAall}),
although its nuclear matter symmetry energy coefficient 
is as large as $a^v_{\rm{sym}}=37.4$ MeV.
For comparison, the corresponding values for IUFSU are
$a_{\rm{sym}} (A=208, T=0)=24.8$ MeV and $a^v_{\rm{sym}}=31.3$ MeV.
In the cases of TM1 and FSU, we obtain $a_{\rm{sym}} (A=208, T=0)=24.3$
and $24.0$ MeV, respectively. As one can see from
the upper-left panel of Fig.~\ref{fig:10EAall},
$a_{\rm{sym}} (A=56)$ obtained in TM1 is almost identical
to that obtained in FSU for $T<3$ MeV, which is the result
of competition between the volume and surface contributions.
From Eq.~(\ref{eq:avas1}), we can see that the surface term
plays a more important role in relatively light nuclei.
Furthermore, we find that the results of $a_{\rm{sym}} (A, T=0)$
for $A=56$, $112$ and $150$ extracted from uncharged nuclei agree
well with those calculated from Eq.~(\ref{eq:avas1}) using parameters
given in Table~\ref{tab:1}. This implies that the mass dependence of
the symmetry energy coefficient of finite nuclei given
by Eq.~(\ref{eq:avas1}) is approximately valid in the relativistic
Thomas-Fermi approximation. From Eq.~(\ref{eq:avas1}),
it is easy to understand the increase of $a_{\rm{sym}} (A, T=0)$
by $\sim$3 MeV from $A=56$ to $A=208$ as shown in Fig.~\ref{fig:10EAall}.
At finite temperature, the symmetry energy coefficient $a_{\rm{sym}}$
decreases smoothly with increasing $T$.
It is found that $a_{\rm{sym}}$ obtained in IUFSU falls more rapidly
than others, which may be related to its lower value of $L$.
It is interesting to compare our results with those of Ref.~\cite{Sub12},
which were obtained in the nonrelativistic Thomas-Fermi approximation.
We find that the temperature dependence of $a_{\rm{sym}}$ shown in
Fig.~\ref{fig:10EAall} is much smaller than that shown in Fig.~5 of Ref.~\cite{Sub12}.
For instance, $a_{\rm{sym}}$ for $A=56$ and $A=112$ with the SBM effective
interaction, as shown by red-dashed lines in Fig.~5 of Ref.~\cite{Sub12},
decreases by $\sim$10 MeV up to $T=8$ MeV, whereas the corresponding
value in our calculation without the Coulomb interaction is $\sim$4 MeV
with the FSU parametrization. From Fig.~\ref{fig:9EATFSU}, one can see
that $a_{\rm{sym}}$ with the Coulomb interaction decreases more rapidly than that without
the Coulomb interaction, and a drop of $\sim$10 MeV is achieved for $A=56$ and $A=112$
with the FSU parametrization, which is very close to the value
of Ref.~\cite{Sub12} with the SBM interaction, as mentioned above.
We note that saturation properties of nuclear matter are very
similar between the SBM interaction and the FSU interaction.
Therefore, the different temperature dependence between our results
shown in Fig.~\ref{fig:10EAall} and that presented in Fig.~5 of Ref.~\cite{Sub12}
may be mainly attributed to different treatments of the Coulomb interaction.
As discussed above, the symmetry energy coefficient of finite nuclei
can be significantly affected by the Coulomb polarization effect.
It would be more appropriate to switch off the Coulomb interaction
in extracting the symmetry energy coefficient of finite nuclei,
rather than to subtract the Coulomb energy after calculating
charged nuclei.

%%%%%%%%%%%%%%%%%%%%%%%%%%%%%%%%%%%%%%%%%%%%%%%%%%%%%%%%%%%%%%%%%%%%%
\section{Conclusions}
\label{sec:4}

We have developed a relativistic Thomas-Fermi approximation by employing
the subtraction procedure for the description of hot nuclei.
The subtraction procedure is necessary for isolating the nucleus
from the surrounding nucleon gas, so that the resulting
properties of the nucleus could be independent of the size of the box 
in which the calculation is performed.
For the nuclear interaction, we have adopted the RMF model and
considered four successful parametrizations, NL3, TM1, FSU, and
IUFSU, which cover a wide range of the symmetry energy slope $L$.
By comparing the results with different RMF parametrizations,
it is possible to study the impact of the RMF parametrization
on properties of hot nuclei.
The correlation between the neutron-skin thickness $\Delta r_{np}$
and the symmetry energy slope $L$ has been confirmed in the present
calculation, and we have found that $\Delta r_{np}$ is almost
independent of temperature $T$. On the other hand, the rms radii
of neutrons and protons, $R_n$ and $R_p$, have been shown to
increase significantly with increasing $T$, and nucleon
distributions become more diffuse at higher temperature.
The excitation energies of hot nuclei have been found to be insensitive
to the RMF parametrization. We have achieved very similar caloric curves
for different RMF parametrizations.

We have investigated the symmetry energy of finite nuclei within
the relativistic Thomas-Fermi approximation.
The symmetry energy coefficient $a_{\rm{sym}}$ of finite nuclei,
which is dependent on the mass number $A$ and temperature $T$, has
been extracted from a nuclear pair ($A$, $Z_1$) and ($A$, $Z_2$)
near the $\beta$-stability line with $Z_1-Z_2=2$.
It has been found that $a_{\rm{sym}}$ calculated from
charged nuclei depends sensitively on the choice of nuclear pair,
especially at high temperature and close to isospin symmetry.
This is because the Coulomb interaction can significantly polarize
nucleon densities and affect nuclear surface properties.
On the other hand, $a_{\rm{sym}}$ calculated from uncharged nuclei
becomes much less sensitive to the choice of nuclear pair in comparison
to that obtained from charged nuclei. Furthermore, the temperature
dependence of $a_{\rm{sym}}$ extracted from uncharged nuclei
has been shown to be much smaller than that calculated
from charged nuclei, which may be due to the Coulomb
polarization effect becoming more pronounced at higher temperature.
Therefore, we conclude that the symmetry energy coefficient of finite nuclei
can be significantly affected by the Coulomb polarization effect.
It would be more appropriate to switch off the Coulomb interaction in
extracting the symmetry energy coefficient of finite nuclei, rather than
to subtract the Coulomb energy after calculating charged nuclei, so that
the resulting $a_{\rm{sym}}$ would not be hindered by the Coulomb polarization.
We have studied the temperature dependence of $a_{\rm{sym}}$
for several representative nuclei using different RMF parametrizations
in order to examine the impact of the RMF parametrization on the
results obtained. It has been found that $a_{\rm{sym}}$ extracted from
uncharged nuclei decreases smoothly with increasing $T$, and a drop of
$\sim$3--$6$ MeV has been achieved up to $T=8$ MeV depending on $A$
and on the RMF parametrization used. The tendency of the temperature
dependence of $a_{\rm{sym}}$ is similar for different RMF parametrizations.
We have compared our results with those obtained in the nonrelativistic
Thomas-Fermi approximation~\cite{Sub12}.
It has been shown that the temperature dependence of $a_{\rm{sym}}$
obtained in the present work from uncharged nuclei is much smaller
than that presented in Ref.~\cite{Sub12}, which may be mainly attributed
to different treatments of the Coulomb interaction.
We have evaluated the ratio of volume and surface symmetry energy
coefficients, $a^v_{\rm{sym}}/a^s_{\rm{sym}}$, at zero temperature.
It has been found that a larger symmetry energy slope $L$ of nuclear matter
corresponds to a bigger value of $a^v_{\rm{sym}}/a^s_{\rm{sym}}$,
which is consistent with that obtained by Skyrme interactions~\cite{SEA09}.
When the temperature increases from zero to a finite value,
both $a^v_{\rm{sym}}$ and $a^s_{\rm{sym}}$ may change with the temperature.
We plan to investigate the temperature dependence of $a^v_{\rm{sym}}$ and
$a^s_{\rm{sym}}$ in a further study.

%%%%%%%%%%%%%%%%%%%%%%%%%%%%%%%%%%%%%%%%%%%%%%%%%%%%%%%%%%%%%%%%%%%%%
\section*{Acknowledgment}

This work was supported in part by the National Natural Science Foundation
of China (Grant No. 11375089).

%%%%%%%%%%%%%%%%%%%%%%%%%%%%%%%%%%%%%%%%%%%%%%%%%%%%%%%%%%%%%%%%%%%%%
%\newpage

%%%%%%%%%%%%%%%%%%%%%%%%%%%
\newpage
%%%%%%%%%%%%%%%
\begin{table}[tbp]
\caption{Parameter sets for the four RMF parametrizations used in this work
and corresponding properties of nuclear matter and finite nuclei.
The masses are all given in MeV.
The quantities $E_0$, $K$, $a^v_{\rm{sym}}$, and $L$ are, respectively, 
the energy per particle, incompressibility coefficient, symmetry
energy coefficient, and symmetry energy slope of nuclear matter
at saturation density $n_0$.
The last two lines show the neutron-skin thickness of $^{208}$Pb,
$\Delta r_{np}$, and the ratio of volume and surface symmetry
energy coefficients, $a^v_{\rm{sym}}/a^s_{\rm{sym}}$, extracted
from the results of $^{208}$Pb at zero temperature.}
\label{tab:1}
\begin{center}
\begin{tabular}{lcccc}
\hline\hline
               & NL3   & TM1   & FSU   & IUFSU \\
\hline
$M$            & 939.0   & 938.0   & 939.0   & 939.0   \\
$m_{\sigma}$   & 508.194 & 511.198 & 491.500 & 491.500 \\
$m_\omega$     & 782.5   & 783.0   & 782.5   & 782.5   \\
$m_\rho$       & 763.0   & 770.0   & 763.0   & 763.0   \\
$g_\sigma$     & 10.2170 & 10.0289 & 10.5924 & 9.9713  \\
$g_\omega$     & 12.8680 & 12.6139 & 14.3020 & 13.0321 \\
$g_\rho$       & 8.9480  & 9.2644  & 11.7673 & 13.5900 \\
$g_{2}$ (fm$^{-1}$) & --10.4310 & --7.2325 & --4.2771 & --8.4929 \\
$g_{3}$        & --28.885 & 0.6183  & 49.8556 & 0.4877  \\
$c_{3}$        & 0.0000  & 71.3075 & 418.3943 & 144.2195  \\
$\Lambda_{\text{v}}$ & 0.000  & 0.000  & 0.030  & 0.046 \\
\hline
$n_0$ (fm$^{-3}$) & 0.148 & 0.145 & 0.148 & 0.155 \\
$E_0$ (MeV)       & --16.3 & --16.3 & --16.3 & --16.4 \\
$K$ (MeV)         & 272   & 281   & 230   & 231 \\
$a^v_{\rm{sym}}$ (MeV) & 37.4  & 36.9  & 32.6  & 31.3  \\
$L$ (MeV)         & 118.2 & 110.8 & 60.5  & 47.2  \\
\hline
$\Delta r_{np}$ (fm) & 0.223 & 0.211 & 0.167 & 0.116 \\
$a^v_{\rm{sym}}/a^s_{\rm{sym}}$ & 3.54 & 3.08 & 2.12 & 1.56 \\
\hline\hline
\end{tabular}%
\end{center}
\end{table}

%%%%%%%%%%%%%%%%%%%%%%%%%%%%%%%%%%%%%%%%%%%%%%%%%%%%%%%%%%%%%%%%%%%%%%%%%%

%%%%%%%%%%%%%%%%%%%%%%%

%%%%%%%%%%%%%%%%%%%%%%%

%%%%%%%%%%%%%%%%%%%%%%%

%%%%%%%%%%%%%%%%%%%%%%%

%%%%%%%%%%%%%%%%%%%%%%%

%%%%%%%%%%%%%%%%%%%%%%%

%%%%%%%%%%%%%%%%%%%%%%%

%%%%%%%%%%%%%%%%%%%%%%%

%%%%%%%%%%%%%%%%%%%%%%%

%%%%%%%%%%%%%%%%%%%%%%%

%%%%%%%%%%%%%%%%%%%%%%%%%%%%%%%%%%%%%%%%%%%%%%%%%%%%%%%%%%%%%%%%%%%%%%%%%%
\end{document}